\documentclass[a4paper,12pt]{article}
\usepackage{graphicx}
\usepackage{amssymb}
\begin{document}

\title{Planar resonant periodic orbits in Kuiper belt dynamics}
\author{George Voyatzis and Thomas Kotoulas\\
University of Thessaloniki, Department of Physics\\
GR-541 24 Thessaloniki, Greece \\email: voyatzis@auth.gr, tkoto@skiathos.physics.auth.gr }

\maketitle

\begin{abstract}
In the framework of the planar restricted three body problem we study a considerable number of resonances associated to the Kuiper Belt dynamics and located between 30 and 48 a.u. Our study is based on the computation of resonant periodic orbits and their stability. Stable periodic orbits are surrounded by regular librations in phase space and in such domains the capture of trans-Neptunian object is possible. All the periodic orbits found are symmetric and there is  evidence for the existence of asymmetric ones only in few cases. In the present work first, second and third order resonances are under consideration. In the planar circular case we found that most of the periodic orbits are stable. The families of periodic orbits are temporarily interrupted by collisions but they continue up to relatively large values of the Jacobi constant and highly eccentric regular motion exists for all cases. In the elliptic problem and for a particular eccentricity value of the primary bodies the periodic orbits are isolated. The corresponding families, where they belong to,  bifurcate from specific periodic orbits of the circular problem and seem to continue up to the rectilinear problem. Both stable and unstable orbits are obtained for each case. In the elliptic problem the unstable orbits found are associated with narrow chaotic domains in phase space. The evolution of the orbits, which are located in such chaotic domains, seems to be practically regular and bounded for long time intervals.              
\end{abstract}

{\bf Keywords.} periodic orbits; Kuiper belt; Elliptic restricted three body problem; 

\section{Introduction}
The dynamics of the long term evolution of small bodies revolving beyond the Neptune's orbit and forming the Edgeworth-Kuiper  belt is a hot research topic in planetary science. Since 1992, when the first small body was discovered  (Jewitt and Luu, 1993), hundreds of such small bodies, called trans-Neptunian objects (TNOs), have been detected and are included in the associated list of the Minor Planet Center (http://cfa-www.harvard.edu/iau/ mpc.html, Jewitt, 1999). The study of their dynamical evolution provides valuable information towards the understanding of the origin, the formation and the structure of our Solar system (Morbidelli et al, 2003). The first theoretical and numerical studies (e.g. Torbett and Smoluchovski, 1990; Kne\v zevi\'c et al, 1991; Duncan et al., 1995; Gallardo and Ferraz-Mello, 1998; see also the review paper by Morbidelli, 1999, and references therein) showed the important role of resonances for the capture and the long term stability of TNO's.     

The most simple but efficient model for understanding the underlying dynamics of the resonant orbits in Kuiper Belt is the restricted three-body problem (RTBP) where the primary bodies are the Sun and Neptune and the small body (of negligible mass) represents a potential TNO. The dynamics can be understood by studying the phase space structure of this simple model, where chaos and order coexist. The simplest model, namely the planar circular RTBP, is a conservative system of two degrees of freedom and for each energy level the particular qualitative characteristics of the dynamics are revealed clearly by calculating the corresponding Poincar\'{e} surface of section. Based on this tool, Malhotra (1996) studied some main mean motion resonances with Neptune and showed that resonances are associated with stable motion which is a libration with respect to the corresponding resonant angle $\sigma=p \lambda ' -q \lambda-(p-q)\varpi$, where $p/q$ denotes the external resonance ($p<q$). This property had been also indicated, using a different approach to the problem, by Morbidelli et al, 1995. The center of libration is a fixed point on the Poincar\'{e} section that corresponds to a resonant stable periodic orbit for the particular energy level.  It is well known that the periodic orbits and their stability are essentially associated with the structure of the phase space (Berry, 1978; Hadjidemetriou,1993,1998). Stable periodic orbits are surrounded by invariant tori and the motion is regular while unstable periodic orbits are associated with a hyperbolic structure. Consequently the usefulness of the computation of families of periodic orbits is to understand the dynamics in an open interval of energy levels without spending an exaggerative computational time. In the planar elliptic problem, which is of two degrees of freedom but non-autonomous, the Poincar\'{e} sections are four dimensional and, subsequently, their usefulness is reduced significantly compared with that of the autonomous case. But the knowledge of periodic orbits provides essential information for the dynamics as in the autonomous case.  It is clarified in Section 2 that such periodic orbits are isolated in a hyperplane of constant eccentricity of Neptune and, thus, information is provided only for a small subset of the phase space. Nevertheless, the localization of a stable periodic orbit in this case is associated with the existence of regular orbits in the neighborhood of such a periodic orbit. Additionally, the numerical simulations show that the eccentricity of Neptune introduce small perturbations to the circular problem that do not affect significantly the long term stability of the resonant librating motion near the stable periodic orbits of the circular problem (Malhotra, 1996; Kotoulas and Voyatzis, 2004a). 

The periodic orbits of the planar circular RTBP have been extensively studied since the decade of '60. A detailed review, which includes the associated definitions, theoretical aspects, references and examples, is given by H\'{e}non (1997). The basic theory and the numerical methods for the computation of periodic orbits in the planar elliptic problem have been described in detail by Broucke (1969a,b) and their application in our Solar system and the asteroidal motion has been shown (e.g. Hadjidemetriou, 1988,1992). Applications also can be found for the dynamics of extra-solar systems (Haghighipour et al, 2003). As far as the Kuiper belt dynamics is concerned, the families of symmetric periodic orbits in the 1/2, 2/3 and 3/4 resonances have been studied (Kotoulas and Hadjidemetriou, 2002) and, as well, the 1/2, 1/3 and 1/4 resonant symmetric and asymmetric families in the planar circular problem (Voyatzis et al 2004). Also, the 2/3 resonant periodic orbits are  studied by Varadi (1999), for various mass ratio and eccentricity values of the primaries.  

The aim of this paper is to study the resonant motion through the computation of the periodic orbits. A considerable number of resonances that are present in the Kuiper belt dynamics and located between 30 and 48 a.u. is included in our study. Namely, we examine the first order resonances $p/q=m/(m+1)$, where $m=1,2,...,7$, the second order resonances 3/5, 5/7 and 7/9 and the third order ones 4/7, 5/8 and 7/10. At this stage we restrict our study to the planar circular and planar elliptic case of the RTBP. In section 2 we present the formulation of the RTBP model and the periodicity conditions used in the present work. In section 3 we present the families of periodic orbits computed for the planar circular model. In section 4 we discuss on the continuation of periodic orbits in the elliptic case and present the results of our computations. Finally, we summarize our results and discuss about the resonant structures obtained and their consequences in the long term stability of the small bodies.

\section{System configuration and periodicity conditions}
We consider the planar restricted three body problem with primaries the Sun and Neptune of mass $1-\mu$ and $\mu$ respectively. The gravitational constant is set equal to unit. In the inertial orthogonal frame $OXY$, the motion of Neptune is either circular or elliptic round the center $O$ with eccentricity $e'$, semimajor axis $a'=1$ and period $T'=2\pi$. In the rotating orthogonal frame of reference $Oxy$, where the Sun and Neptune define the $Ox$-axis,  the motion of the small body is described by the Lagrangian (Roy, 1982) 
\begin{equation}
L=\frac{1}{2}(\dot{x}^2+\dot{y}^2+(x^2+y^2)\dot{\theta}^2+2(x\dot{y}-\dot{x}y)\dot{\theta}) +\frac{1-\mu}{r_1}+\frac{\mu}{r_2},
\label{Lagrangian}
\end{equation}
where $r_1^2=(x+\mu r)^2+y^2$, $r_2^2=(x-1+\mu r)^2+y^2$, $r$ is the distance between the primaries and $\theta$ the angle between the axes $Ox$ and $OX$. For the circular problem it is $r=1$, $\dot{\theta}=1$ and there exist the {\it Jacobi} constant $h$ written in the form  
\begin{equation}
h=\frac{1}{2}(\dot{x}^2+\dot{y}^2-(x^2+y^2))-\frac{1-\mu}{r_1}-\frac{\mu}{r_2}.
\label{Jacobi}
\end{equation}
For the mass of Neptune we use the value $\mu$=5.178$ \times 10^{ - 5}$.

In the circular problem a symmetric periodic orbit can be defined by the initial conditions $x(0)=x_0$, $y(0)=0$, $\dot{x}(0)=0$, and $\dot{y}(0)=\dot{y}_0$ and the periodicity conditions are
\begin{equation}
y(0)=y(T/2)=0, \:\: \dot{x}(0)=\dot{x}(T/2)=0,
\label{PCOND1}
\end{equation}
where $T$ is the period of the orbit which is unknown. We can represent such a periodic orbit as a point in the plane $x_0-h$, where $h$ is the corresponding {\it Jacobi} constant. By varying the value of $x_0$ (or $h$) we get a monoparametric family of periodic orbits with parameter $x_0$ (or $h$). Generally, the period $T$ changes along the family and the eccentricity $e_0$ that corresponds to the initial conditions of the periodic orbit is either almost constant and approximatelly equal to zero or it changes significantly along the family (see Section 2; Henon, 1997). The multiplicity of a periodic orbit is defined as the number of crosses of the orbit with the axis $y=0$ in the same direction ($\dot{y}>0$ or $\dot{y}<0$) in a period. In the elliptic case, the periodic orbits are isolated points in the plane $x_0-\dot{y}_0$ for a particular value of $e'$ and their period is $T=2 k \pi,\:k=1,2,..$. Therefore the same periodicity conditions (\ref{PCOND1}) hold but the value of $T$ is known a priori. A monoparametric family of symmetric periodic orbits is formed by analytic continuation varying $e'$ (the parameter of the family). Such a family can be represented as a curve in the 3D space $x_0-\dot{y}_0-e'$. 

In our computations the periodicity conditions are solved by a Newton-Raphson shooting algorithm (Press et al. 1992) with accuracy $10^{-13}$ or $10^{-11}$ for the circular and the elliptic case respectively. Only for some difficult cases (e.g. near collisions) we were forced to decrease the accuracy by one decimal order. The numerical integrations were performed using a control step Bulirch-Stoer method. The starting points for the computation of the families of periodic orbits in the circular case are obtained from Poincar\'{e} sections at a particular level of the {\it Jacobi} constant. For the elliptic problem we started from the known bifurcation points derived from the families of the circular problem. Our algorithm controls automatically the possible change of the multiplicity of the orbits and the step of the parameter of the family. Since a symmetric periodic orbit crosses the $y=0$ axis at least twice, the algorithm may change the starting point of the periodic orbits when the convergence to the prescribed accuracy fails. The linear stability of the periodic orbits is determined from the corresponding stability indices. For the planar circular case the stability index $k=a_{11}+a_{22}$ is computed, where $a_{ij}$ are the elements of the monodromy matrix (H\'{e}non, 1997). For the planar elliptic case we calculate the indices $k_1$ and $k_2$ introduced by Broucke (1969a,b). For both cases the computation is based on the numerical solution of the variational equations. 

\begin{figure}[ht]
\centering
\includegraphics[width=10cm]{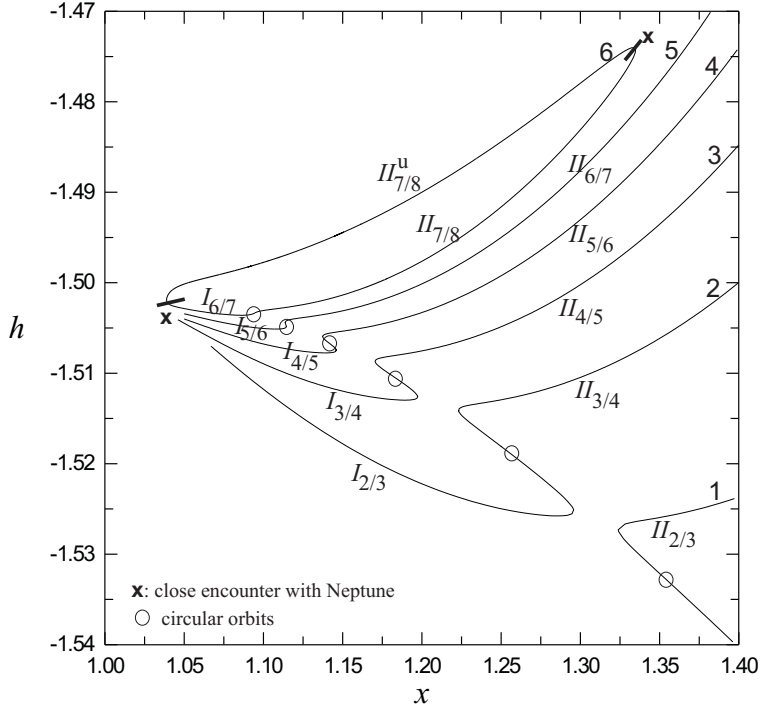}
\caption{Characteristic curves showing the continuation of families of elliptic periodic orbits of first order from the circular ones ($\mu\neq 0$).  The almost straight line segments, indicated by the circles, consist the family of circular orbits (first kind).}  
\label{FF1}
\end{figure}

\section{The planar circular problem}
We consider the planar circular case of the RTBP. The families of symmetric periodic orbits are classified in two different kinds (Poincar\'{e}, 1892; Henon, 1997):\\
{\em Families of circular orbits (or first kind)} : The periodic orbits correspond to nearly circular orbits for the small body. The period $T$, and subsequently the resonance $n/n'=T/T'$, varies along the family.\\
{\em Resonant Families (or second kind)} : The periodic orbits correspond to almost elliptic orbits for the small body. The eccentricity $e_0$ increases along the family but the ratio $n/n'$ is almost constant and rational $n/n'\approx p/q,\: p,q\in Z$.

\begin{figure}[htb]
\centering
\includegraphics[width=14cm]{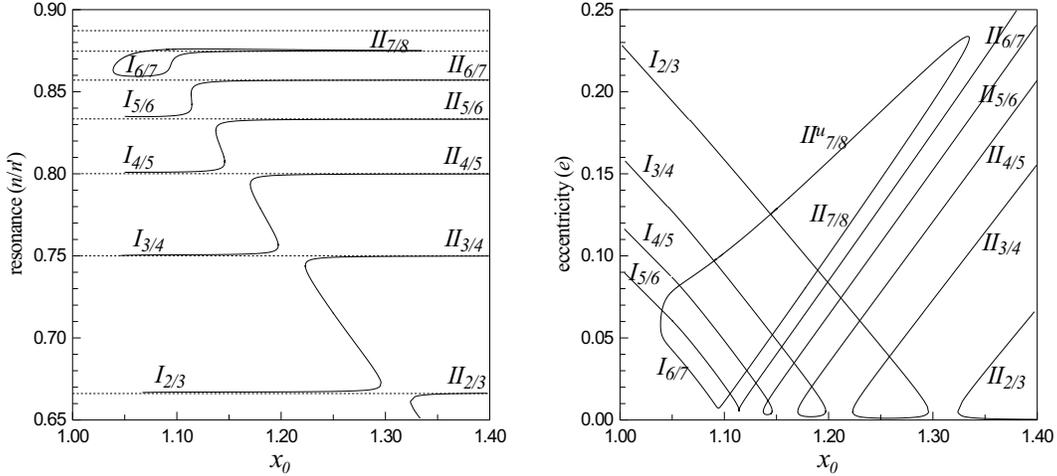}
\caption{Families of resonant periodic orbits for the cases a) 4/5 and b) 5/6. The symbols {\bf x} denote close encounters with Neptune while the encircled ones denote close encounters with the Sun. The segments $I_{4/5}^1$ and $I_{5/6}^1$ are the only that consist of unstable periodic orbits and are presented by a thin curve.}
\label{FF2}
\end{figure}

In the exterior resonances $n/n'=p/q$, studied in this paper, it is $q>p$ and the difference $\Delta=q-p$ defines the order of the resonance. The circular periodic orbits of the unperturbed problem ($\mu=0$) are continued for $\mu>0$ and a family $C$ of first kind exists at each particular value of $\mu$. The families of resonant periodic orbits originate from the $p/q$-resonant orbits of the family $C$. In each resonance, independently of its order, there exist two different families, $I$ and $II$ of second kind, which differ in phase. On family $I$ the small body is initially at perihelion and on family $II$ it is at aphelion. The families may consist of many segments separated by gaps which correspond to close encounter orbits where computations fail. We use the notation $I_{p/q}^n$ (or $II_{p/q}^n$), where $p/q$ denotes the corresponding resonance and $n$ numbers the possible different segments in the family. In the following discussion and plots we select the variable $x_0$ as the parameter for the characteristic curves that present the resonant families.  

\begin{figure}[ht]
\centering
\includegraphics[width=14cm]{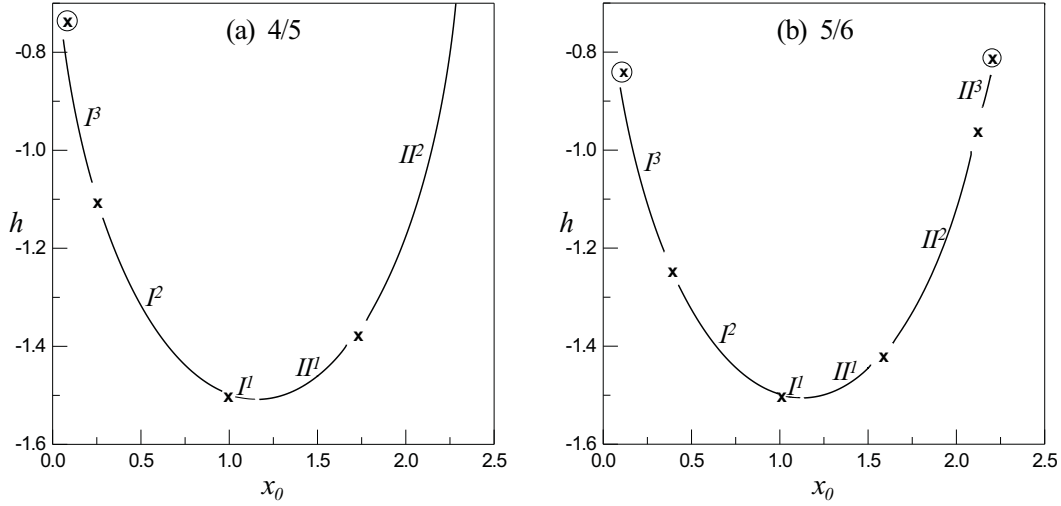}
\caption{The variation of the resonance (left) and the initial eccentricity of the periodic orbits along the families (the parameter of the family is considered to be the variable $x$).}  
\label{FF3}
\end{figure}

\subsection{1st order resonances} 
We consider the first order resonances with $p/q=m/(m+1)$, $m=2,3,..,7$. In this case the circular family of the unperturbed problem breaks close to the resonances when $\mu>0$ and the persisting segments are  separated by gaps. These segments, consisting of almost circular orbits, are continued smoothly giving rise to the resonant families $I$ and $II$. The six characteristic curves computed are presented in Fig. 1. We obtain that a family $I_{m/(m+1)}$ joins smoothly (through a segment of almost circular orbits) with the family $II_{(m+1)/(m+2)}$. By decreasing the parameter $x_0$, the families $I$ approach a collision orbit with Neptune for $h\approx -1.5$. The first order resonances are considered exceptional because they follow the above scenario instead of a regular bifurcation (Guillaume, 1974; Hadjidemetriou, 1993). In our case a rather different structure is observed for the families $I_{6/7}$ and $II_{7/8}$ because the resonant family $I_{6/7}$ avoids the collision with Neptune. A new resonant family of unstable periodic orbits, denoted by $II_{7/8}^u$, is found and the three families ($I_{6/7}$,$II_{7/8}$ and $II_{7/8}^u$) form a closed characteristic curve.     

In Fig. 2a and 2b we present the variation of the resonance and eccentricity, respectively, along the first order resonant families shown in Fig. 1. Actually we present the mean values, computed along one period time interval. We observe that the ratio $n/n'$ vary along the part of the family where the orbits are almost circular. The length of these parts, which corresponds to the the plateau $e\approx 0$ of the curves in Fig. 2b, decreases rapidly as $m$ increases, and disappears for the closed characteristic curve mentioned above.           

In Fig. 1 only the first segment of the families $I_{m/(m+1)}^1$ and $II_{m/(m+1)}^1$ is shown in order to emphasize their origin. In Fig. 3 the complete resonant families 4/5 and 5/6 are presented. For all the resonances with $1\leq m\leq 6$ the family segment $I_{m/(m+1)}^1$ consists of unstable periodic orbits of multiplicity one and terminates at a collision orbit with Neptune. The family continues with new segments separated by collisions with Neptune, and extends up to a collision with the Sun. For $m<4$ (i.e in the resonances 2/3 and 3/4) only one collision orbit is obtained while for $m\geq 4$ two collision orbits exist. In all cases, the segments after the first collision consist of stable periodic orbits.   In the families $II$, the segments $II_{m/(m+1)}^1$ ($1\leq m\leq 7$) reach a collision orbit too, as $h$ increases. Beyond this point the family continues with the segment $II_{m/(m+1)}^2$ which either extends up to a collision orbit with the Sun (cases for $1\leq m\leq 4$) or is interrupted by a second collision orbit with Neptune (cases 5/6 and 6/7). All the families $II_{m/(m+1)}$ consist of stable orbits. Only the periodic orbits which are close encounter orbits prove to be unstable but this result may be an artifact of the limited accuracy of computations. 

\begin{figure}[htb]
\centering
\includegraphics[width=14cm]{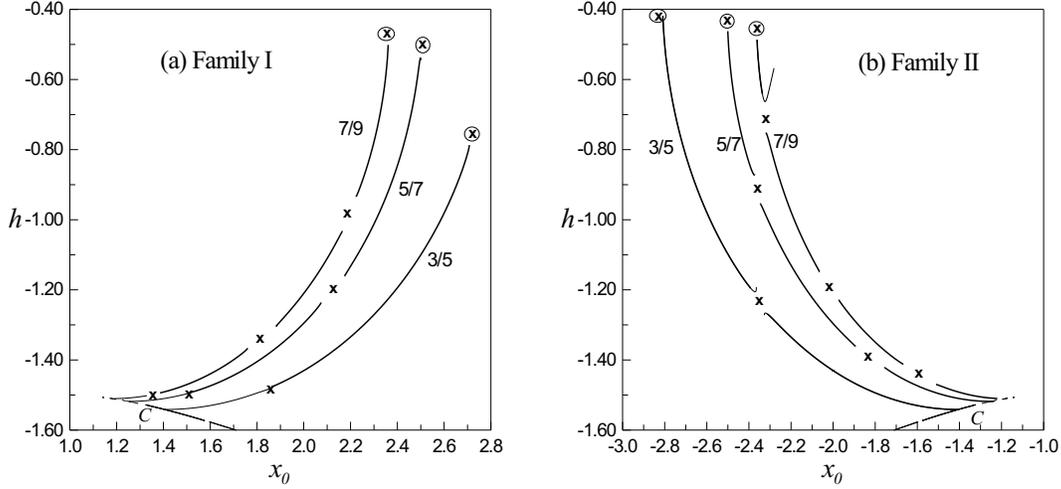}
\caption{Families of periodic orbits for the second order resonances. Bold or thin curves denote stable or unstable orbits respectively. The family $C$ of circular orbits is also indicated. The symbols {\bf x} denote close encounters with Neptune while the encircled ones denote the termination of the family to a collision orbit with the Sun.}
\label{FF4}
\end{figure}

\subsection{Second order resonances}
We consider the second order mean motion resonances $p/q$=3/5, 5/7 and 7/9. The families of resonant periodic orbits bifurcate from the circular family $C$ by analytical continuation for $\mu\neq 0$. The bifurcation points are those on family $C$ where $n/n'=p/q$. For each resonance two families bifurcate, called family $I$ and family $II$. The periodic orbits included in family $I$ cross vertically the axis $Ox$ (i.e. $\dot{x}_0=0$) only for $x>0$, while in family $II$ such a vertical cross occurs only for $x<0$. Thus, the two families are presented in separated plots $x_0-h$ shown in Fig. 4.          

\begin{figure}[htb]
\centering
\includegraphics[width=14cm]{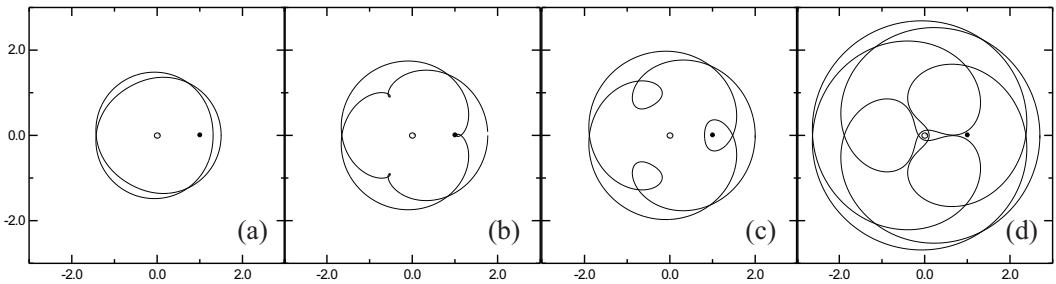}
\caption{Periodic orbits of the family $I_{3/5}$ in the $x-y$ rotating frame. a) $x_0=1.5$, b) $x_0=1.77$ (before the collision with Neptune) c) $x_0=2.0$ (after the collision) and d) $x_0=2.7$ (close to collision with the Sun). The position of the Sun ($x=\mu$) and Neptune ($x=1-\mu$) is indicated.}
\label{FF5}
\end{figure}

The resonant family $I_{p/q}$ starts from the bifurcation point having unstable periodic orbits and is interrupted by a close encounter with Neptune for $h\approx -1.5$. After a close encounter the families continue with stable orbits. In the 3/5 resonance, this continuation extends smoothly up to high eccentricity values (i.e. up to a collision orbit with the Sun). For the 5/7 resonant case, the family reaches to a second close encounter with Neptune and then terminates at a collision orbit with the Sun. In the 7/9 resonant case three collision orbits occur along the family.  The multiplicity of the orbits is affected when a close encounter takes place. An example of some typical 3/5 resonant orbits of the family $I$ is shown in Fig. 5. 

Excluding close encounter orbits, family $II_{p/q}$ consists of stable periodic orbits. Along the families we obtain one, two and three close encounters with Neptune for the 3/5, 5/7 and 7/9 resonances respectively. All families terminate at a collision orbit with the Sun (Fig. 4b). 
   
\begin{figure}[htb]
\centering
\includegraphics[width=14cm]{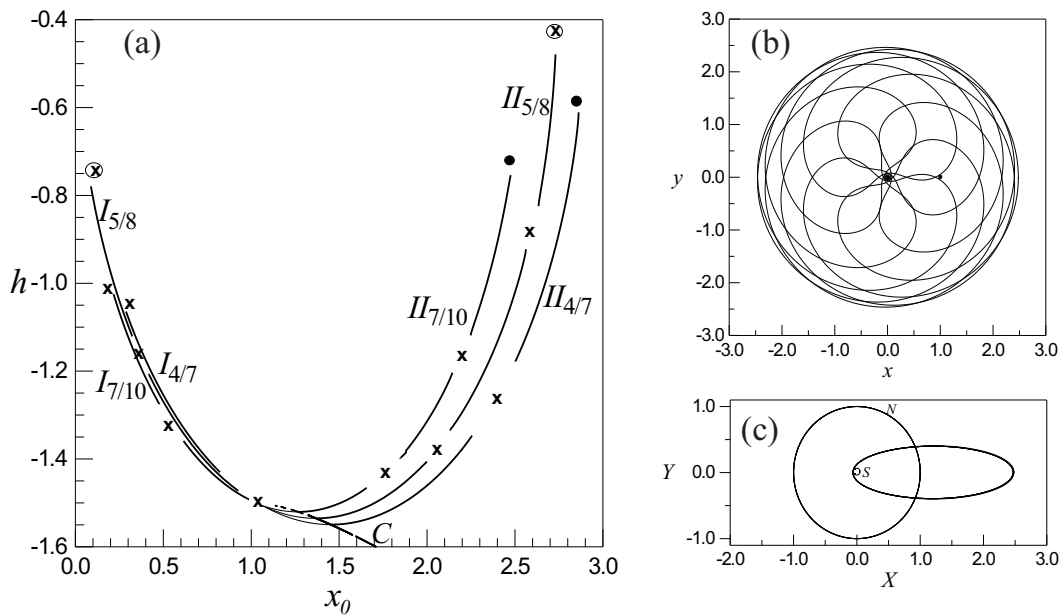}
\caption{a) Families of third order resonances 4/7, 5/8 and 7/10. The symbol {\bf x} and the encircled {\bf x} indicate collisions with Neptune and the Sun respectively. The solid circles indicate almost double collisions. b) A stable periodic orbit of the family $II_{7/10}$ close to a double collision ($h\approx-0.75$) in the rotating $x-y$ frame. c) The same orbit as in (b) presented in the inertial frame X-Y. The circular orbit $N$ of Neptune is also plotted. }
\label{FF6}
\end{figure}

\subsection{Third order resonances}
Our study includes the resonant cases $p/q$=4/7, 5/8 and 7/10. The dynamics of third order resonances shows similar qualitative characteristics to these of the second order. The families $I$ and $II$ of elliptic resonant periodic orbits bifurcate from the circular family and they are presented in Fig. 6a. Families $I$ start with unstable orbits and extend up to close encounter with Neptune at $h\approx-1.50$. The families continue after this point with stable orbits and extend up to a second close encounter. At this point the family $I_{4/7}$ seems to terminate or become strongly chaotic and is difficult to be localized by the computations. This is also the case for the family $I_{7/10}$ after the third close encounter while the continuation of the family $I_{5/8}$ is possible up to a collision orbit with the Sun. 

The periodic orbits of families $II_{p/q}$ are all stable (except in the neighborhood of collisions). Along the families $II_{5/8}$ and $II_{7/10}$ two close encounters with Neptune occur but only one for the family $II_{4/7}$. The family $II_{5/8}$ is continued up to very high eccentricity values and terminates at a collision with the Sun. The families $II_{4/7}$ and $II_{7/10}$ extend up to high eccentricity values and their computation terminates when close encounters with both Neptune and Sun take place. A linearly stable periodic orbit of the family $II_{7/10}$ that approaches the above double collision is shown in Fig.6b and 6c in the rotating and inertial frame respectively. The intersection of the orbits, which is shown in Fig. 6c, is a usual feature of resonant motion. However, because of resonance, there is phase protection and collisions are avoided (Hadjidemetriou 1988, Morbidelli, 1999).

\section{The elliptic restricted three-body problem}
The resonant periodic orbits of the circular problem are bifurcation points (BPs) or, in different terminology,  generating orbits  for the elliptic problem ($e'\neq 0$) when their period is multiple of $T'=2\pi$. Namely, starting from a generating orbit at $e'=0$, a family of periodic orbits of constant period is formed by varying $e'$ (the parameter of the family). In Table 1 we present the values of eccentricity and {\it Jacobi} constant that correspond to the bifurcation points found in the studied resonances (see also Kotoulas and Voyatzis, 2004b). From each BP$l$ two families bifurcate, denoted as $E_{lp}^{p/q}$ or $E_{la}^{p/q}$, where $p/q$ is the resonance, $l$ is the bifurcation point (according to the numbering in the first row of Table 1, which is based on the ascending sorting of the corresponding eccentricity values) and the second subscript $p$ or $a$ denotes the initial position of Neptune is at perihelion or aphelion respectively. The case $n=0$ is discussed in section 4.2. In Table 1, the symbol ``S'' or ``U'' denotes the stability of the bifurcating periodic orbits at $e'\approx 0$ (stable or unstable respectively). The first symbol refers to family $E_{lp}^{p/q}$ and the second one to family $E_{la}^{p/q}$. In all cases, the indicated stability is preserved along the family up to $e'=0.01$, which is the actual eccentricity of Neptune. But we calculate the families and their stability up to high values of $e'$ in order to provide some general results on the resonant structures in the elliptic RTBP. In this case Neptune refers to a fictitious planet.   

\begin{table}[htb]
\begin{center}
\caption{The bifurcation points (BP$l$) where the families $E_{lp}^{p/q}$ and $E_{la}^{p/q}$ of periodic orbits originate. We present the corresponding initial eccentricity value $e_0$ and the Jacobi constant $h$ in brackets. The symbol ``S'' or ``U'' denotes the stability character of the periodic orbits close to the bifurcation point.}  
\begin{tabular}{cccccc}
\hline
$n/n'$ & ~~BP0 & ~~BP1 & ~~BP2 & ~~BP3 & ~~BP4 \\
\hline 
  2/3  & -          & ~~0.469 SU  & -          & -          &  -\\
       & -          & ~~(-1.393)  & -          & -          &  -\\
  3/4  & -          & ~~0.329 SU  & -          & -          &  -\\
       & -          & ~~(-1.452)  & -          & -          &  -\\
  4/5  & -          & ~~0.253 SU  & ~~0.871 SU & -          &  -\\
       & -          & ~~(-1.473)  & ~~(-0.960) & -          &  -\\
  5/6  & -          & ~~0.205 SU  & ~~0.749 SU & -          &  -\\
       & -          & ~~(-1.483)  & ~~(-1.146) & -          &  -\\
  6/7  & -          & ~~0.172 UU  & ~~0.649 SU & ~~0.960 SU &  -\\
       & -          & ~~(-1.488)  & ~~(-1.253) & ~~(-0.743) &  -\\
  3/5  & ~~0.0 UU   & ~~0.427 US  & ~~0.800 SU & -          &  -\\
       & ~~(-1.541) & ~~(-1.428)  & ~~(-1.065) & -          &  -\\
  5/7  & ~~0.0 UU   & ~~0.278 SU  & ~~0.562 SU & ~~0.778 US & ~~0.936 SU \\
       & ~~(-1.518) & ~~(-1.474)  & ~~(-1.325) & ~~(-1.102) & ~~(-0.792) \\
  7/9  & ~~0.0 UU   & ~~0.203 SU  & ~~0.427 SU & ~~0.606 US & ~~0.766 SU \\
       & ~~(-1.510) & ~~(-1.487)  & ~~(-1.406) & ~~(-1.287) & ~~(-1.122) \\
  4/7  & ~~0.0 UU   & ~~0.027 UU  & ~~0.400 UU & ~~0.900 SU &  -\\
       & ~~(-1.549) & ~~(-1.549)  & ~~(-1.408) & ~~(-0.864) &  -\\
  5/8  & ~~0.0 UU   & ~~0.029 SU  & ~~0.335 US & ~~0.800 SU &  -\\
       & ~~(-1.535) & ~~(-1.535)  & ~~(-1.468) & ~~(-1.067) &  -\\
  7/10 & ~~0.0 UU   & ~~0.025 SU  & ~~0.249 US & ~~0.905 SU &  -\\
       & ~~(-1.520) & ~~(-1.520)  & ~~(-1.485) & ~~(-0.872) &  -\\
\hline
\end{tabular}
\end{center}
\end{table}

\begin{figure}[ht]
\centering
\includegraphics[width=14cm]{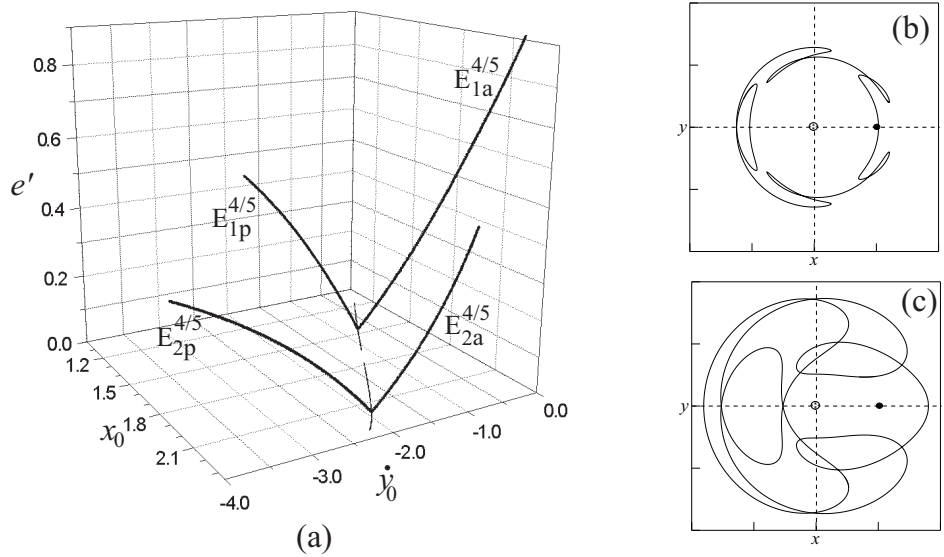}
\caption{a) The resonant families, with $e'$ as a parameter, of the elliptic problem at the resonance 4/5. The families bifurcate from the family $II_{4/5}$ located at $e'=0$ and presented by the thin curve.  b) The stable periodic orbit of the family $E_{1p}^{4/5}$ for $e'=0.4$ in the rotating frame O$xy$. c) the same as in (b) for the family $E_{1a}^{4/5}$. This orbit is unstable.}
\label{FF7}
\end{figure}

\begin{figure}[ht]
\centering
\includegraphics[width=11cm]{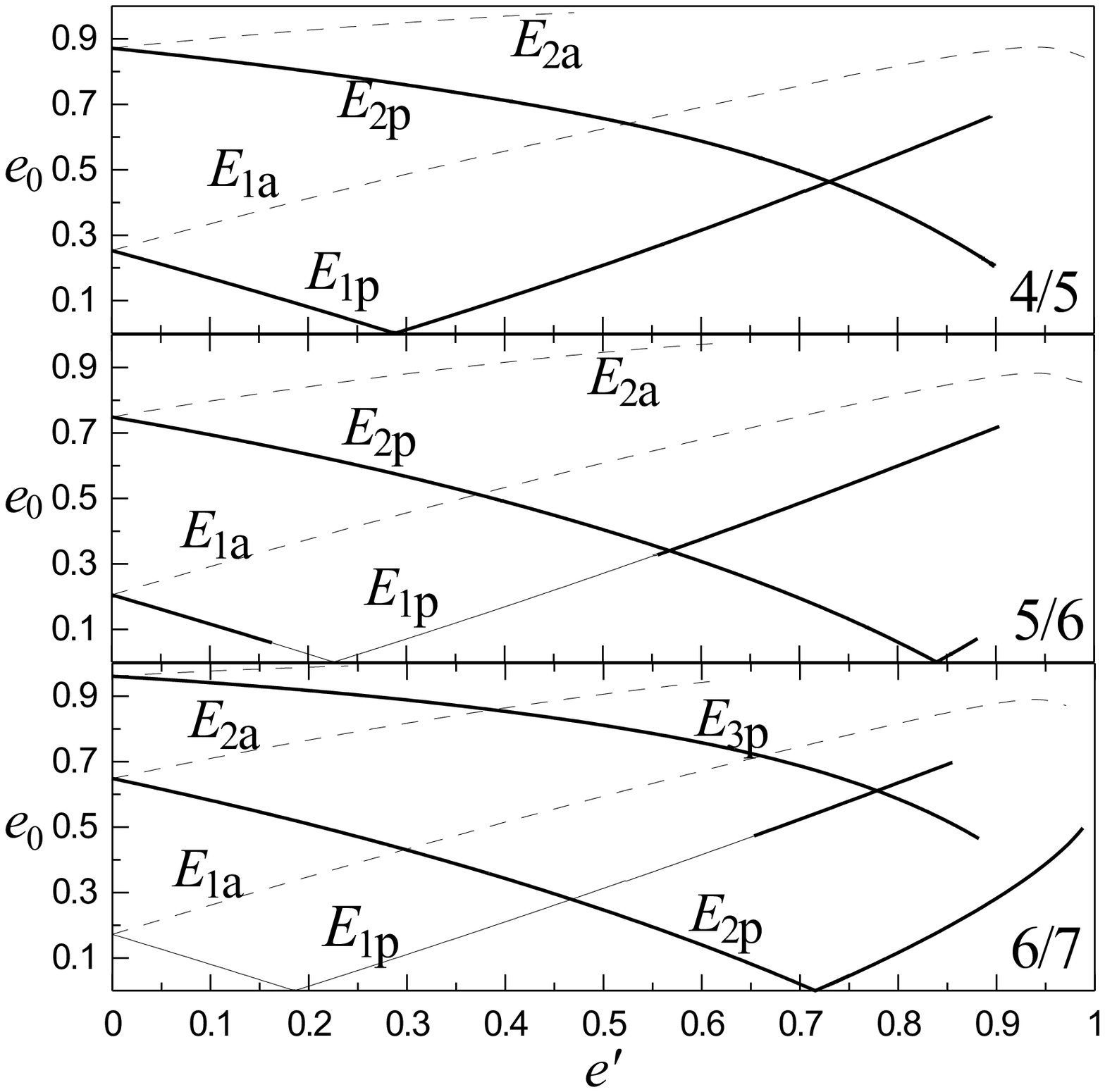}
\caption{The presentation of the families $E_{lp}^{p/q}$ (solid curves) and $E_{la}^{p/q}$ (dashed curves) of first order resonances in the plane $e'-e_0$, where $e_0=e(0)$. Bold curves indicate the family segments of stable periodic orbits and thin curves indicates the segments of unstable ones.} 
\label{FF8}
\end{figure}

\subsection{First order resonances}
The computations reveal that there exist bifurcation points (BPs) of periodic orbits in the elliptic problem, only in the families $II_{p/q}$, ($p/q=m/(m+1)$, $m=1,...,6$). Each one of the 2/3 and 3/4 resonances has one BP and two families originate from each one. These families were studied in Kotoulas and Hadjidemetriou (2002). In the 4/5 resonance there exist two bifurcation points. The families $E^{4/5}_{1p}$ and $E^{4/5}_{1a}$ start from BP1 and the $E^{4/5}_{2p}$ and $E^{4/5}_{2a}$ start from BP2. The families $E^{4/5}_{1p}$ and $E^{4/5}_{2p}$ include stable orbits, while the orbits of the families $E^{4/5}_{1a}$ and $E^{4/5}_{2a}$ are unstable. These families are presented in the $x_0-\dot{y}_0-e'$ space in Fig. 7a. In the same figure the corresponding family of the circular problem is also indicated in the plane $e'=0$. Two samples of periodic orbits of the families $E^{4/5}_{1p}$ and $E^{4/5}_{1a}$ for $e'=0.4$ are shown in Fig. 7b,c. Without loss of important information, it is more convenient to present the families in the projection space $e'-e_0$, where $e_0=e(0)$ is the eccentricity value of the orbit that corresponds to the initial conditions of the periodic orbit. Certainly, a point in the plane $e'-e_0$ does not represent a unique orbit. The variation of the eccentricity $e=e(t)$ along a periodic orbit is rather insignificant and evidently periodic. Therefore, we consider the value $e_0$ as the eccentricity of the whole periodic orbit. By using the above mentioned plane, the computed families are shown in Fig. 8. Solid and dashed curves refer to the families $E_{lp}^{p/q}$ and $E_{la}^{p/q}$ respectively. Also the type of stability is indicated. Thin lines denote stable orbits and bold ones denote unstable orbits.    

In Fig. 8 we observe that in all resonances and along the families $E_{na}^{p/q},\: n=1,2,..$, the initial eccentricity ($e_0$)  increases,  as $e'$ increases, monotonically. The families seem to continue up to the rectilinear case ($e'=1$).  But in most cases and for $e'>0.9$ our computations fail to localize the periodic orbits for various reasons (e.g. the $\dot{y}$ or its derivatives with respect to $y$ or $x$ become large).  Exceptional cases are the families $E^{4/5}_{2a}$, $E^{5/6}_{2a}$ and $E^{6/7}_{3a}$, which start from the BP with the largest eccentricity value. These families tend to terminate at a collision orbit with the Sun as $e'$ increases. For the families $E_{np}^{p/q},\: n=1,2,..$ we observe that they start with a decreasing $e_0$ as $e'$ increases and the orbits are almost elliptic with longitude of pericenter $\tilde{\omega}=180^o$. If the orbits attain the value $e_0=0$, at some particular value $e'$, the family continues with orbits of $\tilde{\omega}=0^o$ and the initial eccentricity increases along the rest of the family. In all families, the stability type is preserved except in families $E^{5/6}_{1p}$ and $E^{6/7}_{1p}$. Especially, in family $E^{5/6}_{1p}$ we observe a segment of unstable orbits in the interval $0.17<e'<0.55$. The end points of such an interval are candidate bifurcation points for families of asymmetric orbits (Beaug\'{e} et al,2003; Voyatzis et al 2004).   

\begin{figure}[ht]
\centering
\includegraphics[width=14cm]{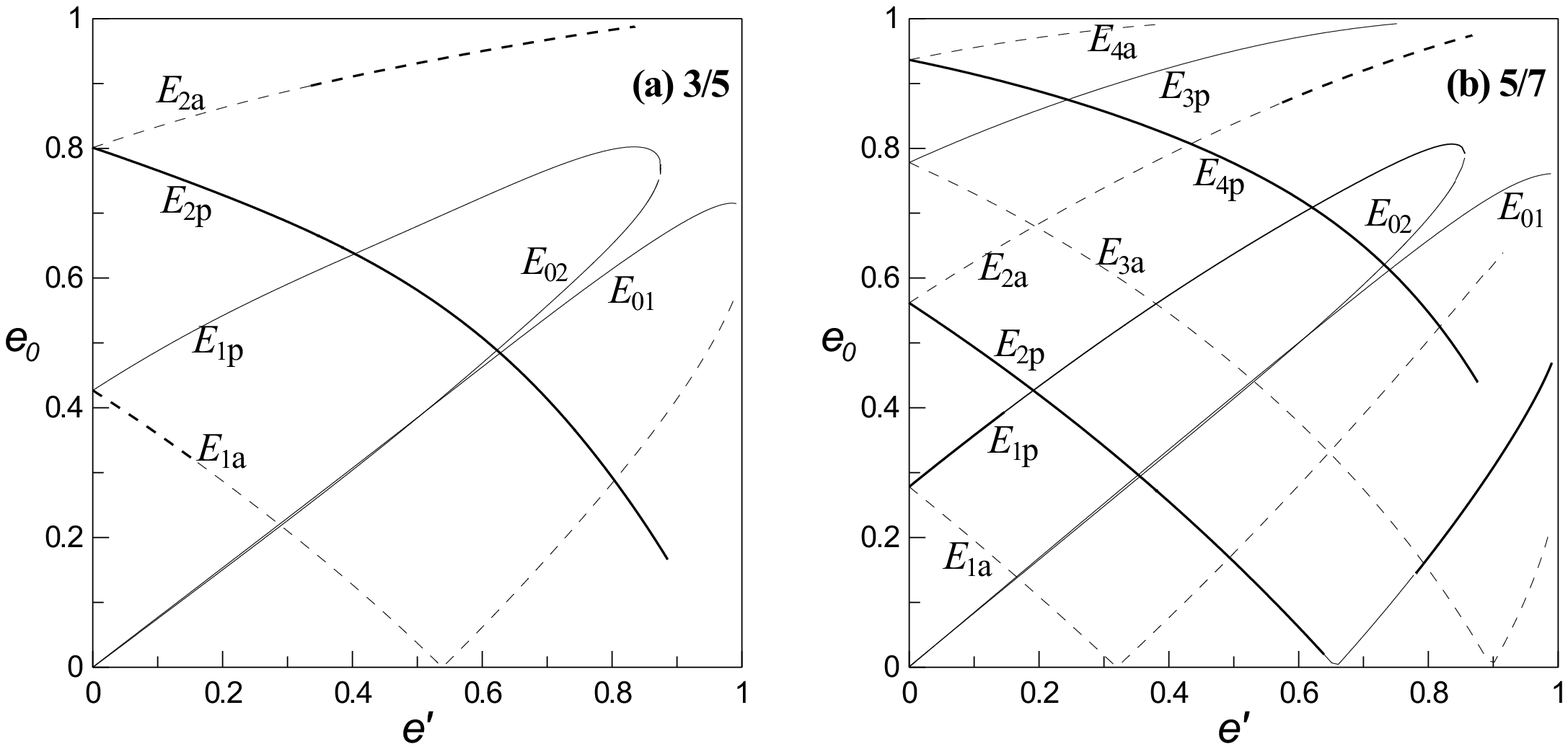}
\caption{Families of periodic orbits of the 3/5 and 5/7 resonances. The presentation is the same as that in Fig. 8.} 
\label{FF9}
\end{figure}

\subsection{Second order resonances}

In the second order resonances ($p/q$, $q=p+2$) both families $I$ and $II$ of the circular problem have bifurcation points, which are starting points for families of the elliptic problem. Additionally, bifurcation points (denoted as BP0 in table 1) exist in the circular family $C$ where the period is $T=q \pi$. Such circular orbits are continued for $e'\neq 0$, providing periodic orbits of period $T=2q\pi$ (Hadjidemetriou 1992). There are two different initial configurations giving rise to these families denoted as $E_{01}^{p/q}$ and $E_{02}^{p/q}$ :\\
Family $E_{01}^{p/q}$ : SUN-$N_{per}$-$B_{per} \:\: (t=0) \:\:\rightarrow$  SUN-$N_{ap}$-$B_{ap} \:\: (t=T/2)$ \\
Family $E_{02}^{p/q}$ : $B_{ap}$-SUN-$N_{per}   \:\: (t=0)\:\:\rightarrow$  $B_{per}$-SUN-$N_{ap} \:\: (t=T/2)$ \\
where $B$ and $N$ denote the small body and Neptune respectively and the subscripts $ap$ and $per$ denote the position of the bodies (at apocenter or pericenter respectively).  

In the case of the 3/5 mean motion resonance there exist three bifurcation points (see Table 1). From BP0, which belongs to the family $C$, the families $E_{01}^{3/5}$ and $E_{02}^{3/5}$ bifurcate with unstable orbits.
The BP1 belongs to the family $II_{3/5}$ of the circular problem and BP2 belongs to a stable segment of the $I_{3/5}$ one. In Fig. 9a the corresponding families are presented in the $e'-e_0$ plane and their stability is indicated. As in the case of first order resonances, along the families $E_{la}^{p/q}$ or $E_{la}^{p/q}$ that bifurcate from the family $II$ the eccentricity $e_0$ increases or decreases, respectively, as $e'$ increases. The opposite situation holds for the families that bifurcate from the $I$ family of the circular problem. In comparison to first order resonances, a different structure is observed for the families $E_{02}^{3/5}$ and $E_{1p}^{3/5}$, which join smoothly at $e'\approx 0.85$ and can be considered as one family starting from BP0 and ending at BP1.  
 
In the case of 5/7 resonance, there exist five bifurcation points. The bifurcation point BP0 belongs to an unstable segment of the family $C$ of the circular problem. The points BP1 and BP3 belong to the family $II_{5/7}$ and the BP2 and BP4 belong to the $I_{5/7}$ one. The corresponding families and their stability is shown in Fig. 9b. As in the case of 3/5 resonance, the families $E_{02}$ and $E_{1p}$ join smoothly. The type of stability of orbits changes twice along the family $E^{5/7}_{2p}$ and a segment of unstable orbits exist. The edges of this segment may be bifurcation points for families of asymmetric orbits, as mentioned above for the $E_{1p}^{5/6}$ family.

The 7/9 resonance shows five bifurcation points distributed as in the 5/7 resonant case. Also, the generated families have the same qualitative features as the ones presented for the 5/7 resonance.

\begin{figure}[ht]
\centering
\includegraphics[width=14cm]{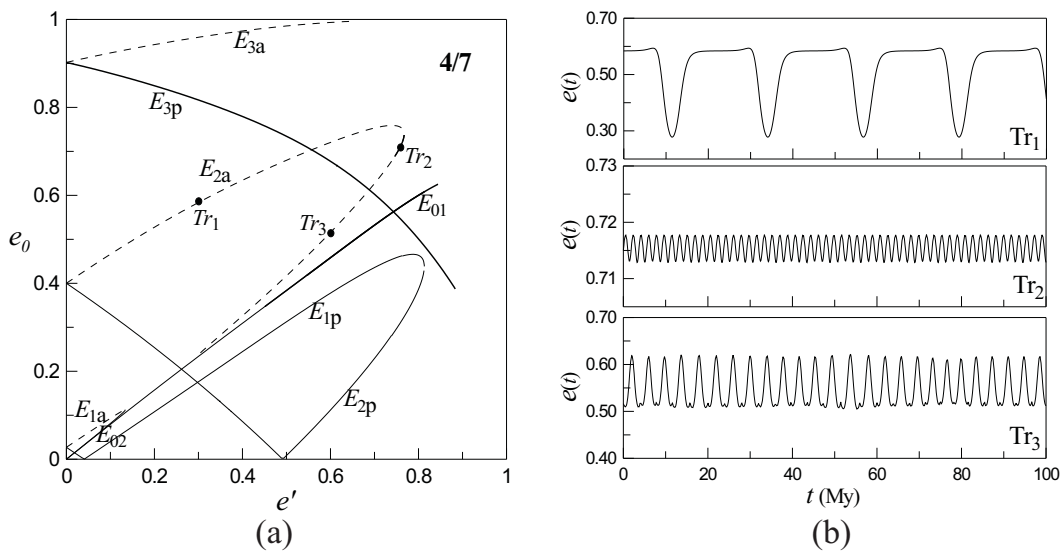}
\caption{a) Families of periodic orbits of the 4/7 resonance in the planar elliptic RTBP. b) The evolution of the eccentricity along some orbits which start near the periodic orbits $Tr_1$ (unstable), $Tr_2$ (stable) and $Tr_3$ (doubly unstable) indicated in (a). Note the different scales in the vertical axis.} 
\label{FF10}
\end{figure}

\subsection{Third order resonances }
The families of the circular problem at the resonances $p/q=$4/7, 5/8 and 7/10 exhibit four points from which families of periodic orbits of the elliptic problem bifurcate (see Table 1). In all cases, the bifurcation point BP0 belongs to the family $C$ and corresponds to a periodic orbit of period $2q\pi/3$. This orbit is continued to the elliptic problem if we assume that it is described three times providing periodic orbits of period $T=2q\pi$. The rest of bifurcation points (BP1, BP2 and BP3) belong to the family $II$. The bifurcating families for the 4/7 resonance and their stability type are shown in Fig. 10a. The families $E^{4/7}_{1a}$ and $E^{4/7}_{02}$ join smoothly at a relatively low value of the primary's eccentricity ($e'\approx 0.13$).  Such a property is also obtained for the families $E^{4/7}_{1p}$ and $E^{4/7}_{2p}$, which join at $e'\approx 0.8$. However, the family $E^{4/7}_{2a}$ does not show a typical continuation. It starts out from BP2 with unstable orbits and continues by increasing $e'$ up to the value 0.767. At this point, $e'$ decreases along the family and the periodic orbits are stable. The stable orbits occupy a short segment ($0.745<e'<0.767$) in the family which is followed by a segment with complex unstable orbits ($0.740<e'<0.745$). For $e'<0.74$ the orbits become doubly unstable and the family terminates at $e'\approx 0.296$.
This termination point is not either a collision orbit or a strongly chaotic one and we are not able to interpret it. Also, there is not any obvious reason for the failure of calculations.

Comparatively to the 4/7 resonance, the resonances 5/8 and 7/10 have four bifurcation points too. However, the stability type of the bifurcating families differs and is indicated in Table 1. The continuation of the families shows the same qualitative characteristics as these described above.     

Closing our study on the elliptic problem we should state that in all cases of unstable periodic orbits the stability indices have values slightly larger than 2.0. This fact, beside numerical integrations, suggests that chaos is not present in a sense of practical importance. Nevertheless, the presence of stable and unstable asymptotic manifolds changes the phase space topology. The orbits that start in the neighborhood of unstable periodic orbits show a large variation in their orbital elements in comparison with that shown by the orbits which start near a stable periodic orbit. A typical example of the evolution of the eccentricity along some orbits of the same family $E^{4/7}_{2a}$ but of different stability type is shown in Fig.10b.  The presented trajectories correspond to the initial conditions $x_0, y_0=0$ and $\dot{y}_0$ which coincide with that of the periodic orbits $Tr_1$, $Tr_2$ and $Tr_3$ indicated in Fig.10a while we have set $\dot{x}_0=10^{-3}$. The time scale is based on Neptune's period, namely  $2\pi$t.u.=165 years.  For the trajectory starting near $Tr_2$, which is stable, the evolution of the eccentricity $e=e(t)$ shows small amplitude oscillations around the starting value. In contrast, the evolution of $e(t)$ along the orbits that start near the simply unstable periodic orbit $Tr_1$ and the doubly unstable $Tr_3$ shows remarkable deviation from the initial value. However, for both cases of instability, the evolution seems regular.

It is worth to mention that the orbits along the families $E^{p/q}_{lp}$ intersect the orbit of Neptune in general (as in the case of Fig. 6c) but the collision of the bodies is avoided because of phase protection. For the families $E^{p/q}_{la}$ the intersection of orbits occurs only for orbits belonging in some particular short segments of the families. We did not find any collision orbits along the families studied in the elliptic RTBP.                  

\section{Conclusions}
In this paper we presented families of resonant symmetric periodic orbits obtained for the planar RTBP and considering $\mu=5.175 \cdot 10^{-5}$ (the normalized mass of Neptune). Thus, our study on the circular RTBP is associated to the Kuiper belt dynamics and its most important resonances between 30 and 48 a.u. Particularly we studied systematically the first order resonances 2/3, 3/4, 4/5, 5/6 and 6/7, the second order ones 3/5, 5/7 and 7/9 and the third order ones 4/7, 5/8 and 7/10 for the circular and the elliptic RTBP. Based on well established methods and previous results of H\'{e}non, Broucke and Hadjidemetriou, we determined all the families of the above mentioned resonances and presented their main characteristic such as bifurcation points, stability, continuation and collisions.         

For the circular planar problem we found that all studied resonances have two families of periodic orbits: the family $I$ and the family $II$. Family $II$ consist always of stable periodic orbits while unstable periodic orbits are obtained only in a short segment of family $I$ located between the origin of the family and the first collision point which is met. Also very short segments of unstable orbits are obtained close to collisions but in these cases the computations are ambiguous. The stable orbits constitute centers for resonant librating motion where Trans-Neptunian objects can be captured. The families, though they are interrupted by collisions, extend up to large values of the Jacobi constant preserving their stability i.e. regular highly eccentric motion can be localized for each resonance at least in the level of the RTBP model. For all resonances we located the bifurcation points at $e'=0$ which are starting points for families of periodic orbits for the elliptic problem.  At each one of them a pair of families ($E_a$ and $E_p$) originates. Thus for $e'\approx 0.01$, which is the eccentricity of the Neptune's orbit we obtain isolated periodic orbits, one in each family. The families of the elliptic problem start withm either stable or unstable orbits. The stability type of the bifurcating families is preserved up to the value $e'$ of the Neptune's orbit, but in many cases it changes for larger values of $e'$. The corresponding eigenvalues of the unstable orbits are always very close to the value +1 (i.e. at least one stability index is slightly greater than 2). This situation has been also observed for internal resonances (Hadjidemetriou, 1993) and indicates the existence of  weak instability at the particular regions in phase space. Indeed, our numerical simulations show that in the neighborhood of unstable periodic orbits of the elliptic problem the motion seems regular for long term evolution but the orbital elements show slow and large variation. Generally, collisions are responsible for the generation of strong chaos but in the elliptic problem all the families found avoid collisions with Neptune. Only the families $E_a$ which bifurcate from the most eccentric bifurcation point seem to terminate at a collision with the Sun as $e'\rightarrow 1$.

As far as the planar RTBP is concerned, our systematic study on resonant periodic orbits verifies the dynamical structures obtained in the past for other resonant cases. Additionally, we should remark the following points: a) The scenario of bifurcation and continuation of the families of first order resonances (Guillaume, 1974; Hadjidemetriou, 1993), shows some differentiation when the resonant families avoid collisions with the planet. In our study, this is obtained for the $I_{6/7}$ and $II_{7/8}$ families, which join together forming a closed characteristic curve b)The families of the elliptic problem seem to extend either up to the rectilinear case ($e'=1$) or join smoothly with another family forming characteristic curves that start from one bifurcation point and terminate to another one. c) It has been suggested (Voyatzis et al, 2004)  that the existence of asymmetric periodic orbits is associated with the existence of continuous segments of unstable orbits along a family of symmetric orbits. This condition is true for the resonances of the form $1/q,\:q=1,2,..$. In our study of the circular problem no any families with such a property have been found. In the elliptic problem the families $E_{1p}^{5/6}$ and $E_{2p}^{5/7}$ include a segment of unstable periodic orbits and, subsequently, we may conjecture the existence of asymmetric periodic orbits in these cases. 

An interesting extension of the present work is to determine the families of periodic orbits in the three dimensional (3D) RTBP. The bifurcation points of families of 3D periodic orbits are given in Kotoulas and Voyatzis (2004b). The 3D resonant structure should provide useful information about the capture of trans-Neptunian objects in highly inclined orbits. Interesting structures of resonant motion can also be provided by the study of asymmetric resonances located beyond 48 a.u. and are not included in the present study.
                                   
\vspace{0.5cm}
\textbf{Acknowledgements}
The authors would like to thank Prof. Hadjidemetriou for fruitful discussions and suggestions. This work has been supported by the research program ``EPEAEK II, PYTHAGORAS, No.21878'' of the Greek Ministry of Education and European Union.

\end{document}